\documentclass[aps,prl,superscriptaddress,twocolumn,preprintnumbers,amsmath,amssymb]{revtex4-1}
\usepackage{graphicx}
\usepackage{dcolumn}
\usepackage{bm}
\usepackage{graphics,graphicx}
\usepackage{epsfig,epstopdf}
\begin{document}

\title{Direct visualization of phase-locking of large Josephson junction arrays by surface electromagnetic waves}
\author{M.\,A. Galin}
\affiliation{Institute for Physics of Microstructures RAS, 603950 Nizhny Novgorod, Russia} \affiliation{Moscow Institute of Physics and Technology (State University),
Dolgoprudny, Moscow region, 141700 Russia}
\author{F. Rudau}
\affiliation{Physikalisches Institut and Center for Quantum Science (CQ) in LISA$^+$, Universit\"at T\"ubingen, 72076 T\"ubingen, Germany}
\author{E.\,A. Borodianskyi}
\affiliation{Department of Physics, Stockholm University, AlbaNova University Center, SE-10691 Stockholm, Sweden}
\author{V.\,V. Kurin}
\affiliation{Institute for Physics of Microstructures RAS, 603950 Nizhny Novgorod, Russia}
\author{D. Koelle}
\affiliation{Physikalisches Institut and Center for Quantum Science (CQ) in LISA$^+$, Universit\"at T\"ubingen, 72076 T\"ubingen, Germany}
\author{R. Kleiner}
\affiliation{Physikalisches Institut and Center for Quantum Science (CQ) in LISA$^+$, Universit\"at T\"ubingen, 72076 T\"ubingen, Germany}
\author{V.\,M. Krasnov}
\email{vladimir.krasnov@fysik.su.se} \affiliation{Department of
Physics, Stockholm University, AlbaNova University Center,
SE-10691 Stockholm, Sweden} \affiliation{Moscow Institute of
Physics and Technology (State University), Dolgoprudny, Moscow
region, 141700 Russia}
\author{A.\,M. Klushin}
\affiliation{Institute for Physics of Microstructures RAS, 603950 Nizhny Novgorod, Russia}

\date{\today}

\begin{abstract}
Phase-locking of oscillators leads to superradiant amplification
of the emission power. This is particularly important for
development of THz sources, which suffer from low emission
efficacy. In this work we study large Josephson junction arrays
containing several thousands of Nb-based junctions. Using
low-temperature scanning laser microscopy we observe that at
certain bias conditions two-dimensional standing-wave patterns are
formed, manifesting global synchronization of the arrays. Analysis
of standing waves indicates that they are formed by surface
plasmon type electromagnetic waves propagating at the
electrode/substrate interface. Thus we demonstrate that surface
waves provide an effective mechanism for long-range coupling and
phase-locking of large junction arrays.
\end{abstract}

 \maketitle


The creation of tunable, monochromatic, high-power and compact sources of electromagnetic (EM) waves in the 0.1-10 THz frequency range remains a serious technological challenge, colloquially known as the
``THz gap" \cite{Tonouchi_2007}. Josephson junctions (JJs) have a unique ability to generate EM radiation with tunable frequency $f=2eV/h$, where $h$ is the Planck constant, $2e$ is the charge of
Cooper pairs and $V$ is the dc-voltage across the JJs \cite{Koshelets_2000,Ozyuzer,Welp,Benseman_2013,Ji_2014,Kadowaki}. The record tunability range, 1-11 THz, \cite{Borodianskyi} has been reported
for intrinsic JJs in cuprate high-temperature superconductors (HTSC), for which the energy gap, determining the upper frequency limit, can be in excess of 20 THz \cite{Doping,Katterwe_Polariton_2011}.
Emission power from a single JJ is small \cite{Koshelets_2000}. It can be amplified in the superradiant manner by phase-locking of many JJs
\cite{Barbara_1999,Ozyuzer,Klushin,Welp,Benseman_2013,Galin_2015,Galin_2018}. However, with an increasing number of JJs their synchronization becomes progressively more difficult due to a rapidly
growing number of degrees of freedom.

Synchronization of large JJ arrays requires long-range interaction
between JJs. Usually it is mediated by resonant cavity modes
leading to formation of standing waves either inside
\cite{Krasnov_2010,Krasnov_2011}, or outside
\cite{Barbara_1999,Klushin} the JJs.
With increasing array size cavity modes get damped by dissipation.
For very large arrays, with sizes significantly larger than the
wavelength of emitted radiation, an alternative, nonresonant
mechanism of synchronization by traveling waves has been suggested
and synchronization of up to 9000 JJs has been demonstrated
recently \cite{Galin_2018}. Traveling waves in JJ arrays are
essentially surface EM waves (SEMWs) propagating at
electrode-substrate or vacuum interfaces. There is a great variety
of SEMWs at metal/insulator interfaces, for a review see e.g.,
Ref. \cite{Sarkar_2017}, including surface plasmons in the
infrared range, which are being actively studied due to
perspectives of the development of plasmonic components for
optoelectronic devices
\cite{Economou_1969,Burke_1986,Ozbay_2006,Zwiller_2010,Zayats_2012,
Zalevsky, PlasmonicsRoadmap_2018,Vlasko_2005}. SEMWs also exist in
superconducting wires \cite{Camarota_2001} and thin films
\cite{Ngai}. Most interesting in the context of this work are
leaky SEMWs \cite{Burke_1986} that facilitate emission of EM power
into open space.

In this work we study Nb-based JJ arrays containing 1500 and 1660 JJs. We employ low-temperature scanning laser microscopy (LTSLM) for visualization of wave dynamics in the arrays. We observe that at
certain bias voltages, corresponding to Josephson frequencies in the sub-THz range, two-dimensional standing-wave patterns appear in the arrays. Our analysis indicates that the standing waves
represent interference patterns of leaky surface plasmon-type surface waves propagating in opposite directions along the electrode/substrate interface. The leakage of SEMW energy into open space
facilitates both emission of EM waves and a long-range interaction between junctions in the array, which is needed for mutual phase-locking and superradiant emission.

\begin{figure}[t]
    \centering
    \includegraphics[width=0.5\textwidth]{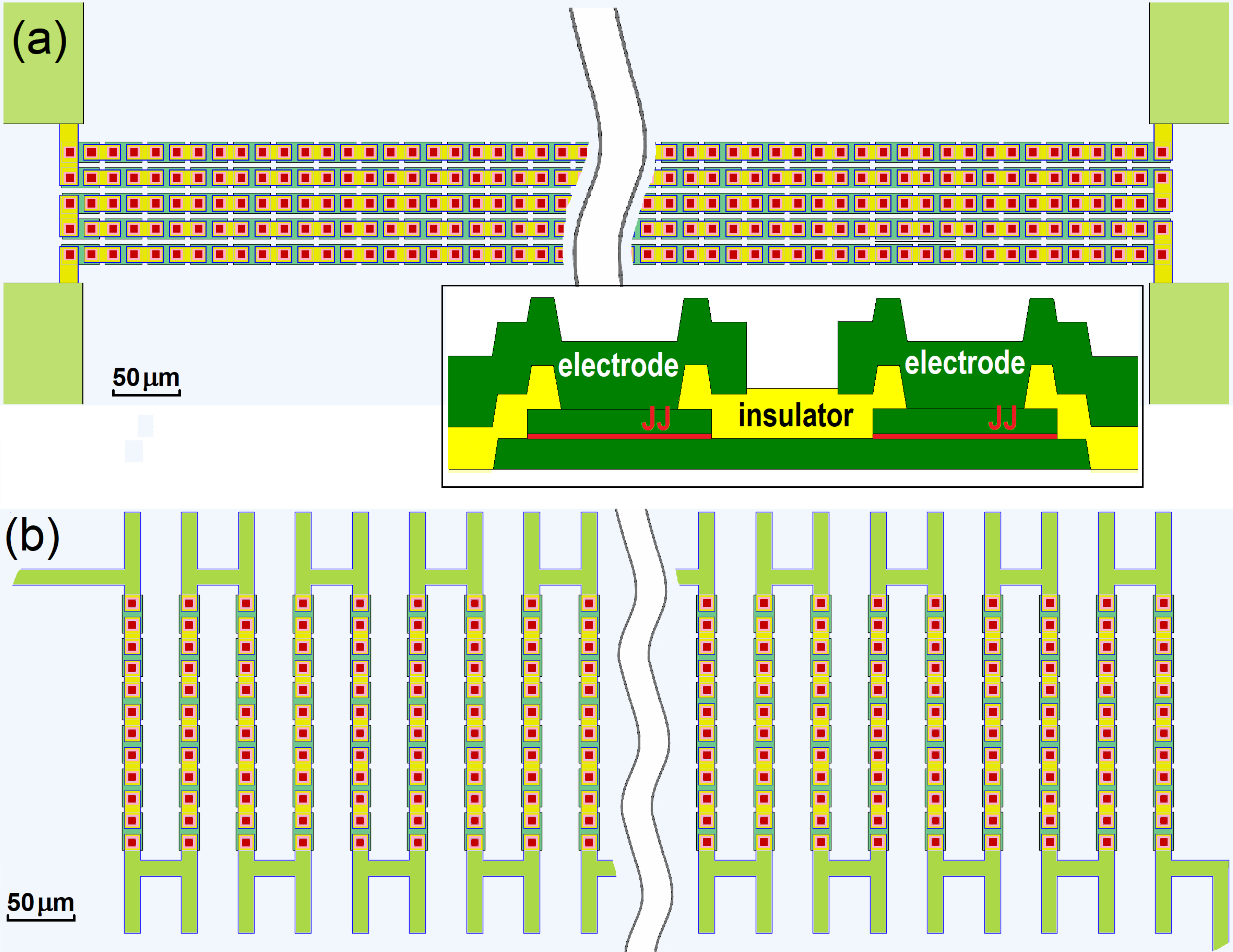}
    \caption{Geometry of studied JJ arrays. (a) Top view of the linear array. Red dots represent junctions, yellow and green stripes are Nb electrodes sequentially connecting the JJs, green rectangles
    at the corners are contact pads. The inset shows a schematic crossection (side-view) of the junction area (adopted from Ref. \cite{Mueller}). (b) Top view of the meander array. Contact pads are outside the image area.
    The vertical wavy lines in (a) and (b) indicate a break in the pictures.
    }
    \label{arrays}
\end{figure}

We study arrays of serially connected Nb/Nb$_x$Si$_{1-x}$/Nb ($x$\,$\sim$\,0.1) JJs.
Fig.\,\ref{arrays} shows layouts of two studied arrays, which we
refer to as ``linear" (a) and ``meander" (b) arrays. JJs with
sizes 8\,$\times$\,8\,$\mu$m$^2$ and a period of repetition of
15\,$\mu$m are formed at the overlap between top and bottom Nb
electrodes, as sketched in the inset. The linear array,
Fig.\,\ref{arrays}(a), consists of five long parallel lines,
containing 332 JJs each, thus yielding in total $N_l$ = 1660 JJs.
The meander array, Fig.\,\ref{arrays}(b), consists of 125
transverse strips with the length $290~\mu$m.
The distance between strips is $40~\mu$m. Each strip contains 12
JJs yielding $N_m$\,=\,1500 in total. The overall size of both
arrays is 5 mm (from left to right in Fig.\,\ref{arrays}). More
details about fabrication and characterization can be found in
Refs. \cite{Mueller,Olaya_2010}. Transport properties and emission
characteristics of similar arrays can be found in Refs.
\cite{Galin_2015, Galin_2018} and in the Supplementary \cite{Supplem}.

We use LTSLM combined with transport measurements for the analysis
of electric field distribution in the arrays. Previously it was
demonstrated that LTSLM and a similar low-temperature scanning
electron microscopy can be used for visualization of standing EM
waves in JJs \cite{Mayer, Kruelle, Quenter,Gerber} and JJ arrays
\cite{Kleiner_2009, Laub, Keck_1998,Lanchenmann}.
In LTSLM a focused laser beam is scanning over the sample surface
causing local heating $\Delta T\lesssim 1$~K, which is small
enough not to destroy superconductivity, but large enough to
induce measurable changes in current-voltage characteristics
(IVC's). The LTSLM image is acquired by applying a certain bias
current through the array and measuring the beam-induced voltage
response $\Delta U(x,y)$ upon scanning of the laser beam in the
($x$,\,$y$) plane. Depending on bias conditions and the
temperature dependence of IVC's, $\Delta U$ can be positive or
negative. $\Delta U>0$ is due to a suppression of the critical
current $I_c$ and the switching of the junction under the beam
spot from the superconducting to the resistive state. $\Delta U<0$
is due to the reduction of the quasiparticle resistance of the
junction heated by the beam. More details about LTSLM and the
measurement setup can be found in the Supplementary
\cite{Supplem}.

\begin{figure*}[tbh]
    \centering
    \includegraphics[width=0.97\textwidth]{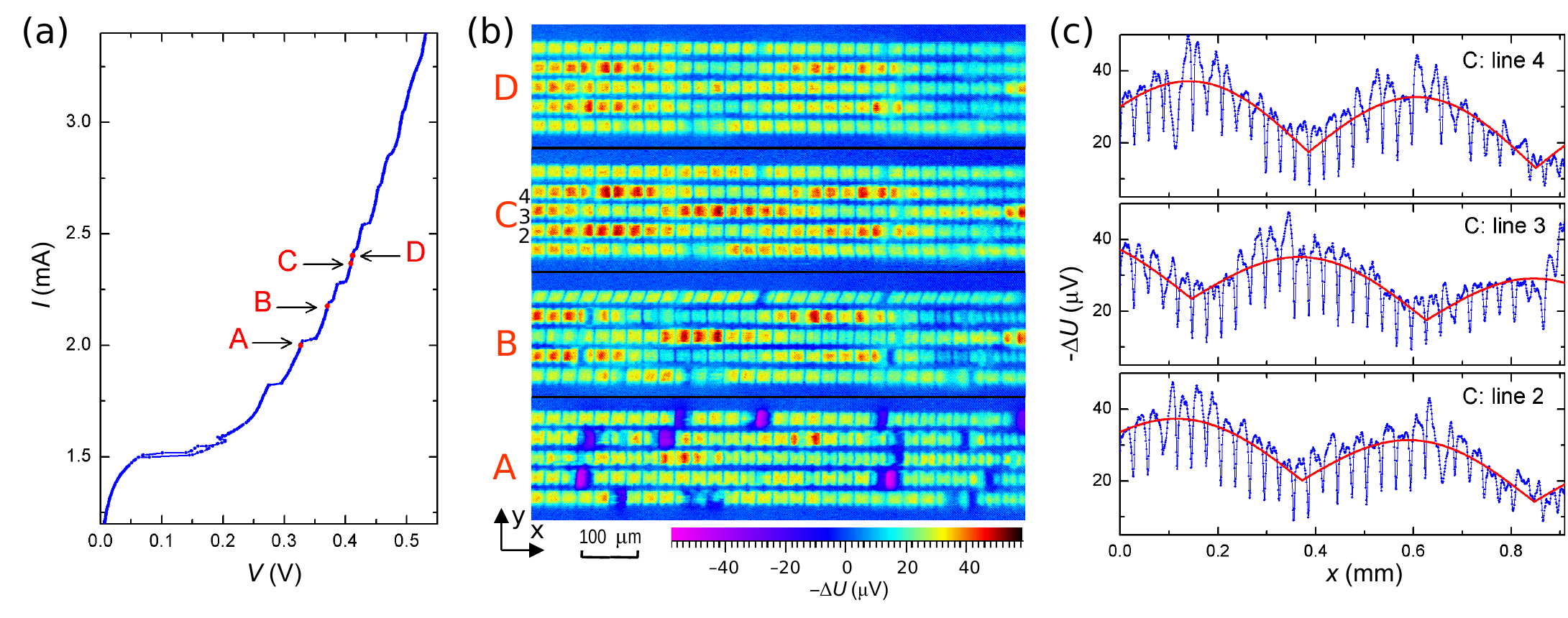}
    \caption{LTSLM analysis of the linear JJ array at $T\sim 6$~K. (a) $I$-$V$ characteristic of the array measured during LTSLM imaging.
    (b) LTSLM images obtained at four bias points, indicated in (a). The length of scans along $x$-axis is $L_x\,=\,0.91$~mm. The scans are stretched about 2 times along $y$-axis for better viewing.
    Development of standing-wave correlations is clearly seen in patterns B, C and D. (c) LTSLM response (blue lines) along horizontal array lines 2\,--\,4 (from bottom to top as indicated in (b),
    panel C) for the pattern C, $I=2.37$~mA. The data are averaged over the width of strip. Red lines represent fitting curves obtained by the method of least squares. Antisymmetric modulation in
    neighbor lines is seen.}
    \label{IVC_linear}
\end{figure*}

Fig.\,\ref{IVC_linear}(a) shows the IVC of the linear array
recorded during the LTSLM measurements at the base temperature
$T\simeq 6$~K. A series of current steps is seen in the IVC, as
reported earlier for similar arrays \cite{Galin_2015, Galin_2018,
Klushin_2011}.
The steps do not occur for a single JJ \cite{Olaya_2010}. They are
caused by wiring of JJ's in the array and, therefore, depend on
electrode geometry. At lower temperature ($T$\,$\sim$\,4\,--\,5~K)
it can be ascertained \cite{Supplem} that steps have a regular
spacing, $\Delta V \simeq 39$ mV, corresponding to a
characteristic frequency $f_r=2e \Delta V/h N_l \simeq 11.4$ GHz.
Such a low frequency may originate only from geometrical
resonances in the long, $L_l\simeq~5$~mm, straight lines of the
array, see Fig. \ref{arrays}(a). The resonant frequency,
$f_r=c/2L_l\sqrt{\varepsilon^*}$, where $c$ is the speed of light
in vacuum, yields the effective dielectric permittivity
$\varepsilon^*\simeq 6.9$ \cite{Galin_2015,Supplem}.

Fig.\,\ref{IVC_linear}(b) represents LTSLM images $\Delta U(x,y)$
of a part of the linear array at four bias currents, marked by
arrows in Fig.\,\ref{IVC_linear} (a). The horizontal scanning
field, $L_x\simeq 0.91$\,mm, corresponds to $\simeq 18\%$ of the
total array length. At the lowest bias point A, $I=2.0$ mA and
$V\simeq 0.33$ V, some junctions (pink and blue spots) are still
in the superconducting state, due to some inhomogeneity of JJs in
the array. Otherwise there is no well defined spatial variation of
the array response. At a slightly higher bias point B, $I=2.17$ mA
and $V\simeq 0.371$ V, a certain alternation with maxima and
minima of $\Delta U(x,y)$ along the $x$-direction appears in the
three middle lines.
At point C, $I=2.37$ mA and $V\simeq 0.409$ V, a clear
standing-wave pattern develops in the whole array. It has an
antisymmetric modulation with maxima in one line corresponding to
minima in the neighbor lines. This is demonstrated in detail in
Fig. \ref{IVC_linear}(c), which represents averaged scans of
$\Delta U(x)$ along each of the three middle lines of the array
(blue) together with fitting curves (red). 
An additional minor increment of the bias current to
point D, $I\simeq 2.4$~mA and $V\simeq 0.412$~V, leads to a
visible reconstruction of the standing-wave pattern. Here it
becomes almost symmetric with aligned maxima and minima in all the
lines.

The increase of voltage 
leads to the growth of the Josephson frequency 
and a reduction of the wavelength of EM radiation.
Indeed, such a tendency can be traced from 
Fig. \ref{IVC_linear}(b). The Josephson frequencies at points B, C
and D are: $f(B)\simeq 107.8$~GHz, $f(C)\simeq 119.0$~GHz and
$f(D)\simeq 119.9$~GHz. The corresponding periods of standing
waves
are $\Delta x(B)=0.53$~mm, $\Delta x(C)=0.47$~mm, and $\Delta
x(D)=0.46$~mm. Since LTSLM probes only the EM amplitude, the EM
wavelength is twice the period of LTSLM image, $\lambda = 2\Delta
x$. The observed decrease of $\lambda$ with increasing $f$ follows
a simple relation $\lambda=c/f\sqrt{\varepsilon^*}$ with
$\varepsilon^* = 6.9 \pm 0.2$, consistent with the estimation
above from the step voltages in the IVC.

The observation of a correlated two-dimensional standing-wave
order indicates a global synchronization of the whole array. 
We did LTSLM scans over a broad bias range
along the IVC. However, such correlated standing wave patterns
were observed only in a limited bias range from slightly below the
point B to slightly above the point D. This is qualitatively
consistent with a narrow bias range in which significant EM wave
emission occurs from such an array \cite{Supplem}.

\begin{figure*}[tbh]
    \centering
    \includegraphics[width=0.97\textwidth]{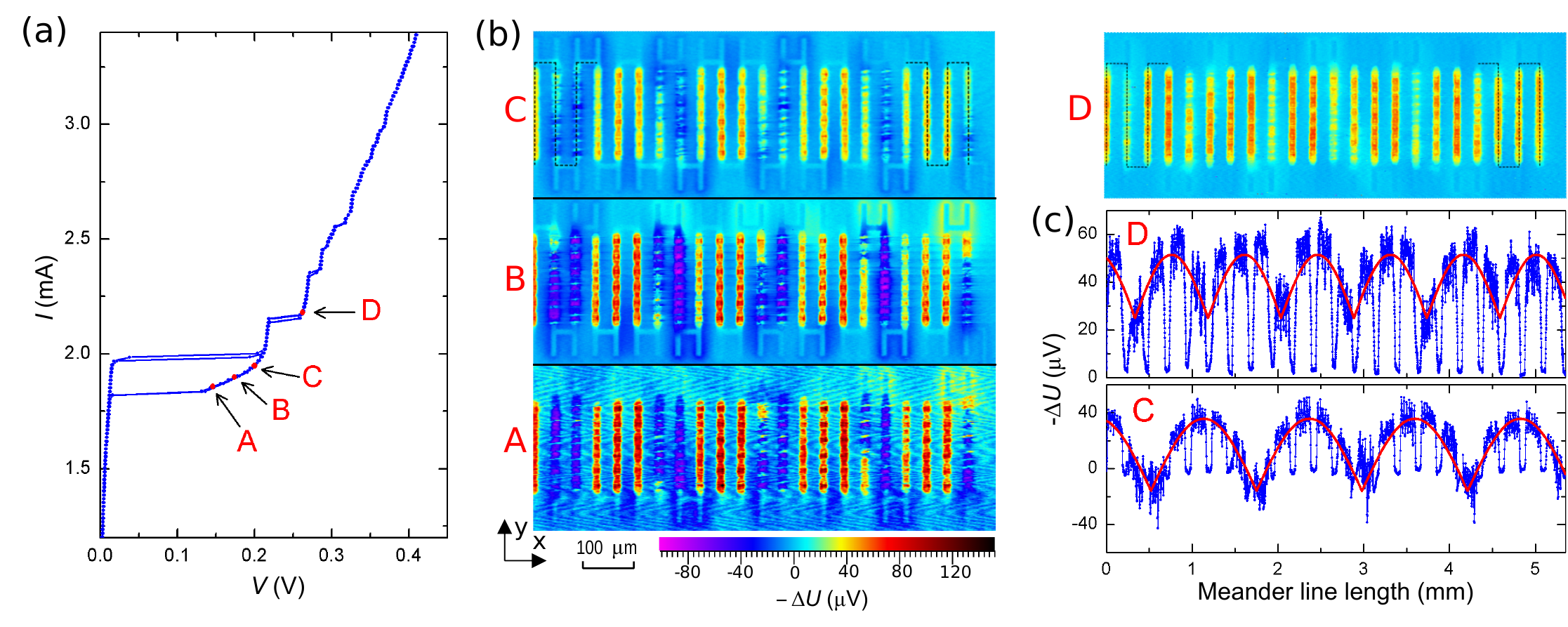}
    \caption{LTSLM analysis of the meander array at $T\sim 5$~K. (a) $I$-$V$ characteristic of the array, measured during LTSLM imaging. (b) LTSLM images at different bias points, indicated in (a).
    The length of scans along $x$-axis is $L_x\,=\,0.91$~mm. Development of standing-wave pattern is clearly seen. Note that the standing wave is not periodic in the horizontal direction. Dot lines
    indicate the start and the end of track along the length of the meander line where the periodicity of response is observed.(c) LTSLM responses (blue lines) along the tracks indicated in (b) from
    the bottom-left to the bottom-right corner of the pattern at the bias points C (top) and D (bottom). The data are averaged over the width of strip. Red line represents fitting curves obtained
    by the method of least squares. Clear periodicity along the whole meander length indicates that standing waves are formed by plasmon-type surface waves propagating along Nb electrodes.}
    \label{IVC_meander}
\end{figure*}

Fig.\,\ref{IVC_meander}(a) shows the IVC of the  meander array
recorded during LTSLM measurements at $T\simeq 5$~K. Similar to
the linear array, the IVC of the meander array also has distinct
resonant steps.
Fig.\,\ref{IVC_meander}(b) represents LTSLM images of the meander
array at different bias points marked in
Fig.\,\ref{IVC_meander}(a). A: $I$\,=\,1.86\,mA and
$V$\,=\,144\,mV, B: $I$\,=\,1.9\,mA and $V$\,=\,174\,mV, C:
$I$\,=\,1.95\,mA and $V$\,=\,200\,mV, D: $I$\,=\,2.18\,mA and
$V\,\simeq\,263$~mV. Note that the IVC of the meander array
exhibits a hysteresis, presumably due to self-heating. Points A--C
are measured at the reverse part of the IVC within the hysteretic
area. LTSLM images are taken at the right end of the array with
the same field of view as in Fig. \ref{IVC_linear} (b), $L_x\simeq
0.91$~mm, which encompasses 22 transverse strips.

Standing wave patterns in the horizontal direction can be seen in
all images of Fig. \ref{IVC_meander} (b). The periodicity $\Delta
x$ is gradually decreasing with increasing voltage from A to D: in
A it is about five transverse strips and in D about three. This is
in a qualitative agreement with the expected decrease of the
wavelength with increasing frequency. However, if we calculate the
speed of EM wave propagation along the $x$-axis, $c_x=2\Delta x
f$, with $f = 2eV/hN_m$, we obtain unreasonably low values. For
example, at bias point A the periodicity along the horizontal axis
is $\Delta x \simeq 200~\mu$m, which yields $c_x \simeq 1.8 \times
10^7$ m/s $\simeq 0.06 c$. This would require a huge dielectric
constant $\varepsilon^*\sim 300$, which does not make sense.
Therefore, the observed standing-waves in the meander array can
not be caused by propagation of a volume EM wave in some media.
This conclusion is also confirmed by the observation that with
changing frequency the nodal areas (blue color) are shifting along
the meander line and may start/end at an arbitrary place of the
transverse strips. E.g. in A they usually start/end at the edges,
but in C - in the middle of the strips. Thus, the standing wave
pattern is not periodic along the horizontal $x$-axis, as would be
expected for a straightforward EM-wave propagation in the
substrate. Those inconsistencies, revealed by a specific meander
geometry of the array, which has a large geometrical deceleration
\cite{Gandolfo}, provide a clue to understanding of the nature of
EM waves in our arrays.

In Fig.\,\ref{IVC_meander}(c) we plot the LTSLM response 
for the image C (top) and D (bottom panel) along the meander line,
as a function of the overall length of the Nb electrode from the
bottom-left to the bottom-right edges of the images in the
Fig.\,\ref{IVC_meander}b. From this plot it becomes clear that
there is a long-range standing wave order {\em along the
electrode}. The periodicity is $\Delta l = 1.23$~mm for the image
C and $\Delta l = 0.85$~mm for the image D. The corresponding
phase velocities, $c^*=2\Delta l f$, are C: $\simeq 1.59 \times
10^8$ m/s and D: $\simeq 1.44 \times 10^8$ m/s, which yield a
reasonable $\varepsilon^*\simeq 4.0 \pm 0.3$ \cite{Note1}. The
performed estimations unambiguously prove that EM waves, building
the standing wave, are propagating along the electrode line and a
significant value of $\varepsilon^*$ indicates that they propagate
at the Nb/substrate side. This is a signature of surface EM waves
at metal/insulator interfaces
\cite{Sarkar_2017,Economou_1969,Burke_1986,Ozbay_2006,Zwiller_2010,Zayats_2012,
Zalevsky,PlasmonicsRoadmap_2018,Vlasko_2005,Camarota_2001,Ngai}.

Another confirmation of SEMW character of standing wave resonances
in the meander array comes from transport measurements. At low
temperatures we observe a very fine step structure in the IVC (see
the Supplementary \cite{Supplem}) with a typical voltage
separation $\Delta V\simeq 7$~mV. It yields a very low primary
resonant frequency $f_r \simeq 2.3$~GHz corresponding to a long
resonator length $\simeq 3$~cm,
which is consistent with the total length of the meandering line.
This confirms that such resonances are formed by traveling surface
waves bound to Nb electrodes.

There are many known modes of SEMWs \cite{Sarkar_2017}. While we can not provide a decisive distinction of the SEMW mode in our case, we argue that the intermediate value of $\varepsilon^*\sim
6.9-4.0$ inbetween Si (11.9) and vacuum (1) suggests that those are leaky surface-plasmon-type SEMWs propagating along one interface and leaking energy at the opposite interface of the metallic film
\cite{Burke_1986}. However, the actual sub-THz frequency is well below the plasma frequency. Consequently they correspond to the linear part of the dispersion relation for surface plasmons (for an
additional discussion see the Supplementary \cite{Supplem}). Importantly, the leaky nature of the involved SEMW both facilitates long-range synchronization of a large array and enables EM wave
emission into open space. All this is a prerequisite for creation of a high-power coherent THz oscillator.

To conclude,
synchronization of large oscillator arrays is a challenging problem. In this work we performed simultaneous transport measurements and low-temperature scanning laser microscopy of large arrays with
1500 and 1660 Josephson junctions. Our main result was the observation of standing-wave patterns, indicating global phase-locking of the arrays. From an analysis of the evolution of standing wave
patterns with changing Josephson frequency we deduced that those patterns are formed by plasmon-type surface EM waves propagating along electrodes at the superconductor/substrate interface. We
conclude that such type of surface waves can facilitate both emission of power and phase-locking of very large oscillator arrays, which is required for creation of high-power THz sources.

\section*{Acknowledgements}
We are grateful to F.\,M\"uller and Th.\,Scheller (PTB
Braunschweig, Germany) for sample fabrication. V.M.K. is grateful
for hospitality during a sabbatical period at MIPT, arranged via
the 5-top-100 program.
The work was supported by the Russian Science Foundation, 
grant No.\,20-42-04415 (experimental part: carrying out the measurements - Figs.\,\ref{IVC_linear}(a), \ref{IVC_linear}(b), \ref{IVC_meander}(a), \ref{IVC_meander}(b)), the Russian Foundation for
Basic Research, grant No.\,18-02-00912 (theoretical part: fitting - Figs.\,\ref{IVC_linear}(c), \ref{IVC_meander}(c)), the State Contract No.\,0035-2019-0021 "Transport properties and electrodynamics
of nanostructural superconductors and hybrid systems: quantum effects and nonequilibrium states", the Deutsche Forschungsgemeinschaft via project KL930/17-1, the COST action NANOCOHYBRI (CA16218),
European Union H2020-WIDESPREAD-05-2017-Twinning
project SPINTECH under Grant Agreement No. 810144
and the Swedish Research Council, project 2018-04848.


\end{document}